\documentclass[10pt,letterpaper]{article}
\usepackage{opex3}
\usepackage{cite}
\begin{document}

\title{Modeled fiber amplifier performance near the mode instability threshold}
\author{Arlee V. Smith$^*$ and Jesse J. Smith}
\address{AS-Photonics, LLC, 8500 Menaul Blvd. NE, Suite B335, Albuquerque, NM, USA 87112}
\email{*arlee.smith@as-photonics.com} 

\begin{abstract}{}
We numerically model fiber amplifier performance near and slightly above the mode instability threshold. These results are compared with recently published experimental work. Using weakly amplitude modulated pump light we obtain qualitative agreement with the measured instability thresholds, mode switching ranges, pixel power modulations, and modal amplitude modulations.  
\end{abstract}
\ocis{(060.2320) Fiber optic amplifiers and oscillators; (140.6810) Thermal effects; (190.4370) Nonlinear optics, fibers; (290.5870) Rayleigh scattering.}

\section{Introduction}

Numerous reports on the performance of high power fiber amplifiers have noted a striking instability in the spatial profile of the output beam. Above a sharp threshold power the signal beam quality degrades\cite{Eidam1,Haarlammert,Stutzki2}, rendering the amplifier unsuitable for coherent beam combining. In previous papers we demonstrated that stimulated thermal Rayleigh scattering can cause this modal instability\cite{SmithSmith1,SmithSmith3}. This stimulated process amplifies a weak, frequency shifted, higher order mode from a low initial power to a level comparable to the power in the fundamental mode. The necessary high gain leads to a sharp power threshold. This gain mechanism requires a small red frequency shift of the higher order mode relative to the fundamental mode. The frequency shift that produces the highest gain typically lies in the range 300 Hz to 3 kHz and depends on the fiber parameters and operating conditions. At powers below the instability threshold the spectrum of the amplified light in the higher order mode is primarily a Stokes shifted version of the light in the fundamental mode. Well above threshold the Stokes shifted light in the higher order mode is strong enough to amplify twice Stokes shifted light in the fundamental mode. Far above threshold this leads to a cascade of frequencies at multiples of the Stokes shift, giving a comb of frequencies for both transverse modes. In addition, high order modes other than LP$_{11}$ can be excited with various frequency shifts. In this paper we use numerical simulations to study the behavior of the modal powers and their spectra at pump powers ranging from slightly below to somewhat above the first instability threshold. 

A recent experimental paper by Otto {\it et al.}\cite{Otto2} provides measurements of mode beating spectra and its evolution with output power in the vicinity of threshold. We will compare our model predictions with their results. The experimental results are for a counter-pumped amplifier. However, modeling the counter-pumped fiber using our current model is prohibitively time consuming, so we model a similar co-pumped fiber instead. The source of difficulty for the counter-pumped amplifier is that our model integrates all optical fields from the signal input end of the fiber to the signal output end. This requires that we specify the pump power at the signal input end. This is not a problem for a co-pumped fiber, but it implies that to match a target modulation level for the applied pump power for a counter-pumped fiber we must precompensate the modulation of the pump at the signal input end. This is an iterative procedure that sometimes requires a large number of model runs. 

Further complicating counter-pumped behavior near and above the instability threshold is a positive feedback effect. The higher order mode is usually less efficient than the fundamental in extracting energy from the pump, particularly if the Yb$^{3+}$ doping is confined to the center portion of the light guiding core. This implies that when the signal light is transferred from the fundamental mode to the higher order mode near the output end of the fiber, the pump light penetrates deeper into the fiber, leading to earlier mode switching. This feedback tends to make the transition from the fundamental mode to the higher mode more abrupt in counter- than in co-pumped fibers. It may also accentuate the exchanges of power between the fundamental and higher order mode that will be described below. 

For these reasons the modeled performance of the co-pumped fiber is not expected to quantitatively match the experimental counter-pumped amplifier. However, we expect a comparison of qualitative similarities to provide a strong test of our proffered mode coupling mechanism and our numerical model. Detailed modeling of the counter-pumped fiber operating above threshold must await further speed up of our numerical model.

\section{Model parameters}

Descriptions of our numerical model were published earlier \cite{SmithSmith1,SmithSmith3}, and we plan to present full mathematical details in a future paper. We use the amplifier parameters listed in Table \ref{tab.table2} throughout this paper. They are chosen to nearly match the fiber labeled LPF45 in \cite{Otto2}, but with a somewhat larger mode field diameter of 68 $\mu$m. As in LPF45, the Yb$^{3+}$ doping is confined to a diameter of 63 $\mu$m which is somewhat smaller than the core diameter of 81 $\mu$m. Other important model parameters include a $64 \times 64$ spatial grid, a 64 point per cycle time grid, and a propagation step length of 6 $\mu$m. We use a step index profile to approximate the photonic crystal guiding of LPF45.
\begin{table}[htb]
\centering\caption{Parameters of test amplifier.}\label{tab.table2}
\begin{tabular}{cc|cc}\hline 
$d_{core}$&81 $\mu$m&$d_{dopant}$ &63 $\mu$m\\
$d_{clad}$&255 $\mu$m&$d_{outer}$&$-$\\
$\lambda_p$&976 nm&$\lambda_s$&1040 nm\\
$\sigma_p^a$&2.47$\times 10^{-24}$ m$^2$&$\sigma_p^e$&2.44$\times 10^{-24}$ m$^2$\\
$\sigma_s^a$&5.8$\times 10^{-27}$ m$^2$&$\sigma_s^e$&5.2$\times 10^{-25}$ m$^2$\\
${P}_{signal}$&see text&${P}_{pump}$&see text\\
$dn/dT$&1.2$\times 10^{-5}$ K$^{-1}$&$L$&1200 mm\\
$n_{core}$&1.45015&$n_{clad}$&1.450\\
$\tau$&901 $\mu$s&$N_{Yb}$&3.0$\times 10^{\;25}$ m$^{-3}$\\
$R_{bend}$&$\infty$&$A_{\rm eff}(LP_{01})$&3600 $\mu$m$^2$\\
beat length ($LP_{01}-LP_{11}$)&29.9 mm&$A_{\rm eff}(LP_{11})$&3420 $\mu$m$^2$\\
scattering loss&$0$&linear absorption&$0$\\
LP$_{01}$, LP$_{11}$, LP$_{11}^{\;\prime}$ linewidths&$0$&$C$&703 J/kg-K\\
$K$&1.38 W/m-K&$\rho$&2201 kg/m$^3$\\
thermal boundary&$202.5\times 202.5$ $\mu$m&boundary condition&$T=300$ K\\\hline
\end{tabular}
\end{table}

To provide a baseline we first model the amplifier when it is seeded by quantum noise, or actually by a simple simulation of quantum noise in which we inject 10$^{-16}$ W of Stokes shifted signal in LP$_{11}$ plus 10 W of unshifted signal in LP$_{01}$. As in all the results presented in this report we use the steady-periodic Green's function method\cite{SmithSmith3}, including harmonics at (0,1,2,3,4) times the Stokes frequency shift, to analyze the thermal response to periodic heating at each spatial pixel. We varied the Stokes shift to find the frequency with the lowest threshold for co- and counter-pumped fibers. For co-pumping the gain is maximum near 600 Hz; for counter-pumping the frequency of maximum gain shifts to near 500 Hz with a threshold power that is 92\% of the co-pumped threshold. The gain line widths are quite broad\cite{SmithSmith3}. The frequencies and threshold powers differ for co- and counter-pumping because the transverse profiles of the upper state population differ, and this leads to different heat distributions.

We analyze the oscillatory output signal field into a time varying set of fiber transverse modes, and each mode is further analyzed into a set of frequencies. For the baseline, quantum seeded amplifier the output spectra of LP$_{01}$ and LP$_{11}$ are found to be simple until power substantially exceeds threshold. Below that they have the same spectra as the injected signals, with a single frequency component for LP$_{01}$ at zero frequency shift, and a single frequency component for LP$_{11}$ shifted by the Stokes frequency. There is no obvious amplitude modulation of the individual modes.

\section{Measured performance}

Key findings of Otto {\it et al.}\cite{Otto2} for fiber LPF45 include the following: above threshold the light is primarily confined to modes LP$_{01}$ and LP$_{11}$, including both orientations of LP$_{11}$; the total amplified signal power is nearly constant in time even above the mode instability threshold; the signal measured by a small area power detector near the center of the signal output beam shows a power spectrum consisting of three narrow harmonic peaks near 350, 700, and 1050 Hz; the signal mode instability threshold is approximately 200 W which should correspond to approximately 220 W of pump if unit quantum efficiency is achieved (the pump power is not presented); the same detector shows a gradual transition from low to high amplitude modulation as the pump power is ramped through the threshold region; well above threshold the spectra of power through an aperture broadens to 2 kHz, and no clear spectral lines are discernable.

\section{Amplitude modulated pump model}

The measured performance detailed in Section 3 is in marked contrast to the modeled baseline performance described in Section 2. The measured threshold was near 200 W which is well below the threhold we compute for quantum level seeding. Further, the observed power spectra of the spatial pixels consisted of sharp lines rather than the broader lines expected from quantum noise seeding. They also found that the power at the center of the output beam was modulated at 350 Hz. All these features are consistent with the use of a pump or signal input that is amplitude modulated at 350 Hz. This would reduce the threshold by an amount that depends on the modulation strength\cite{SmithSmith3}. It would also cause the individual modes to acquire an amplitude modulation at 350 Hz, and it would produce narrow pixel power spectra. 

We think this is strong evidence that the experiment uses an amplitude modulated pump or signal. We will use a pump that is sinusoidally power modulated at 350 Hz with a relative modulation of $\pm 10^{-3}$. The 350 Hz modulation frequency is chosen to match the experimentally observed modulation frequency. The modulation depth of $10^{-3}$ is chosen without knowledge of the actual experimental pump characteristics. It is intended to test whether our model provides qualitative agreement with the observed properties of the experimental fiber. We use an unmodulated injected signal consisting of 9.9 W in LP$_{01}$ and the 0.1 W in LP$_{11}$ with no frequency shift between them. The amplification process impresses the pump modulation pattern onto the signal light, producing weak, symmetric frequency side bands on the signal in both seeded modes, LP$_{01}$ and LP$_{11}$. The Stokes shifted, modulation-induced side band of LP$_{11}$ is preferentially amplified by stimulated thermal Rayleigh scattering, while the anti Stokes sideband is deamplified\cite{SmithSmith3}. Because the power in the Stokes shifted LP$_{11}$ mode exceeds the quantum noise level near the fiber input, the threshold is reduced well below the baseline value.

Because we are making only qualitative comparisons with experiment for reasons listed above, and because our computed threshold powers do not agree closely with the measured values, we choose for simplicity to normalize the computed pump and signal powers to a single arbitrary power. On this scale the instability threshold is near a pump power of 0.5.

\subsection{Modal content versus pump power}

Using the conditions just specified we compute the evolution of modal fractions shown in Fig. 1 as the pump power is ramped through threshold. These modal powers are time averaged over the 350 Hz beat cycle.
\begin{figure}[htb]
\centering\includegraphics[width=11cm]{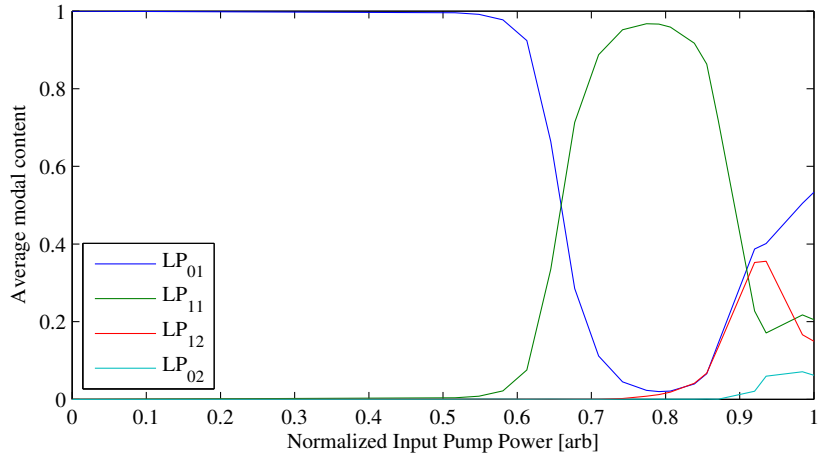}
\caption{\label{fig.dispersive}Evolution of the time averaged modal content versus pump power in the co-pumped fiber. The LP$_{11}$ mode is seeded with 0.1 W of unmodulated light; the pump is sinusoidally amplitude modulated by $\pm$0.1\% at 350 Hz. The instability threshold is at a normalized input pump power of approximately 0.5.}
\end{figure}

\subsection{Periodic modal modulation}

Operating slightly above threshold, the baseline case with an unmodulated pump gave individual modal powers that were nearly constant in time. In contrast, the modulated pump produces modal powers that are amplitude modulated at 350 Hz. This is illustrated in Fig. 2 where we show the time-dependent modal powers. The upper plot is at a pump power of 0.645; the lower plot is at a pump power of 0.71, which is well above threshold. In both plots the total signal power (blue curves) is only slightly modulated, while the individual modes exchange power at 350 Hz.
\begin{figure}[htb]
\centering
\begin{tabular}{c}
\includegraphics[width=11cm]{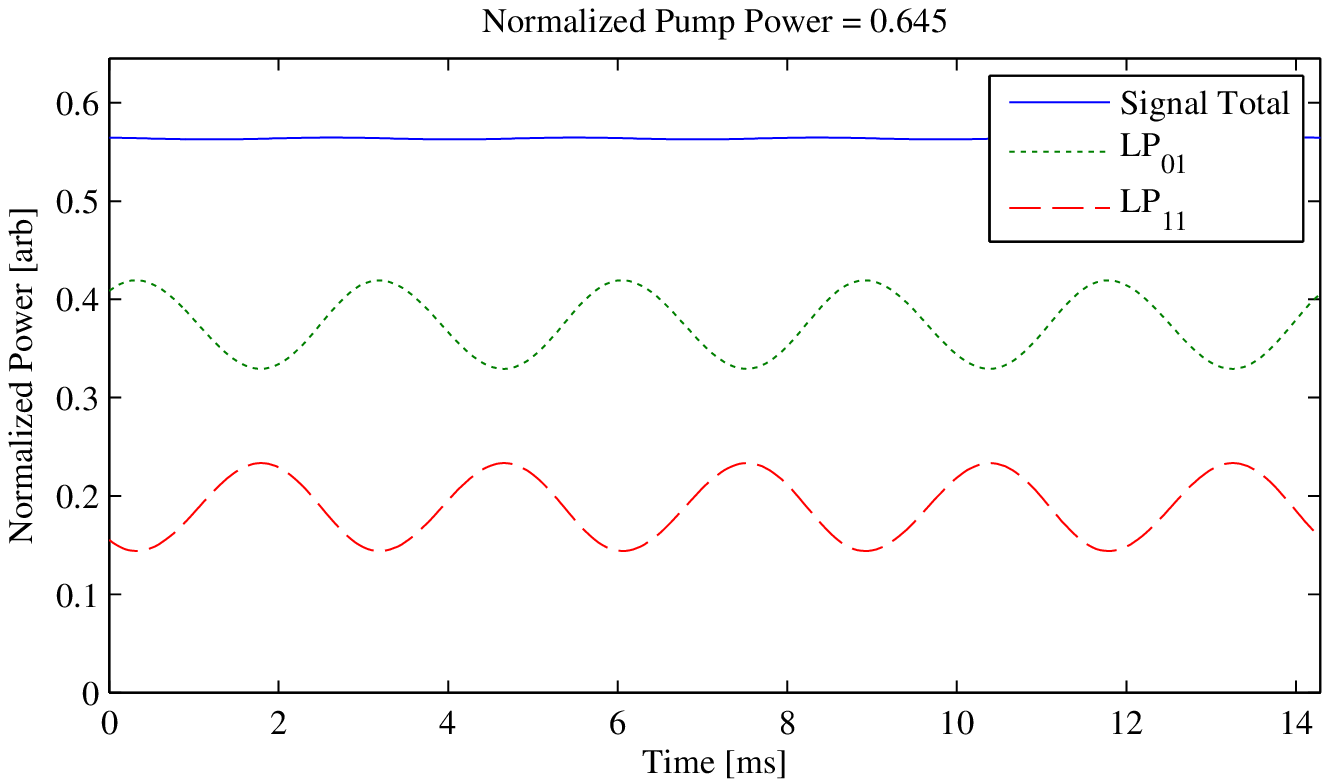}\\
\includegraphics[width=11cm]{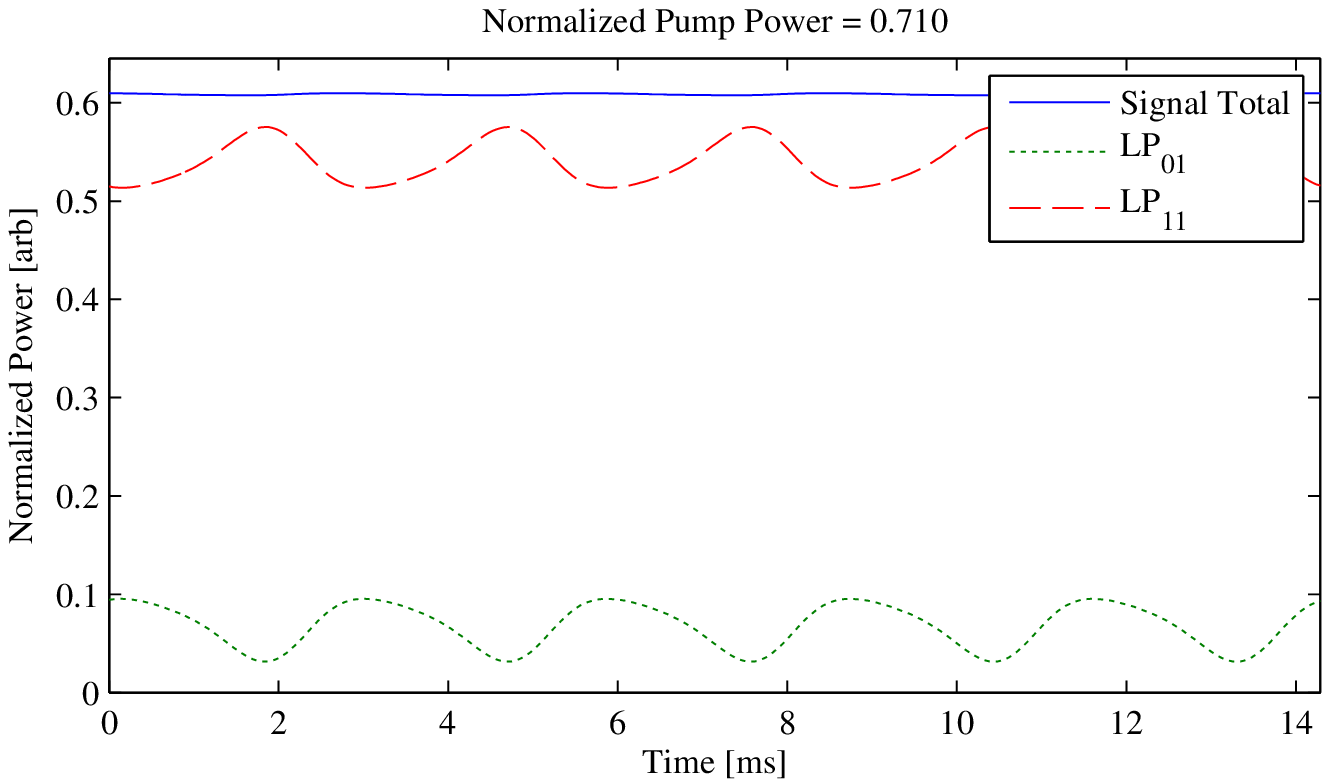}\\
\end{tabular}
\caption{\label{fig.dispersive}Modal content versus time for co-pumped fiber at pump input powers of 0.645 (upper plot) and 0.71 (lower plot). The conditions are the same as in Fig. 1. The individual modes are strongly modulated at 350 Hz, but the total signal power is only slightly modulated.}
\end{figure}

\begin{figure}[htb]
\centering\includegraphics[width=11cm]{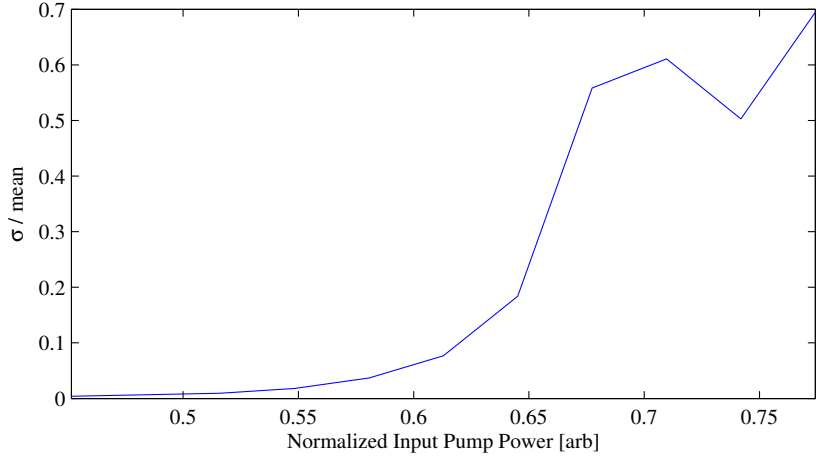}
\caption{\label{fig.modulation}Amplitude modulation at the beam center versus pump power. The vertical axis is the standard deviation of the signal power normalized to the time averaged signal power.}
\end{figure}

For comparison with the measured modulation of the output signal presented in \cite{Otto2}, we analyze the depth of modulation at the beam center as a function of pump power. Figures 3 and 4 of \cite{Otto2} display the amplitude modulation of the signal power as the pump power is swept through the region of the instability threshold. The amplifier studied for those figures was different from LPF45, but presumably LPF45 behaves similarly. In the measurements a small detection aperture was positioned near the beam center so it measured primarily the amplitude modulation of the LP$_{01}$ mode. Their figures show that the standard deviation of the signal power, normalized to the average of the pump power, increases from zero to approximately 0.5 over a power range of 80 to 160 W. We have computed a similar plot of beam center modulation using our model and display it in Fig. \ref{fig.modulation}. It shows how the same measure of modulation used in \cite{Otto2} increases from zero to approximately 0.5 as the pump power increases from 0.45 to 0.75. The simulation result is in good qualitative agreement with the experimental result, at least up to a normalized power of 0.7. 

\subsection{Pixel modulation spectra}

\begin{figure}[htb]
\centering\includegraphics[width=11cm]{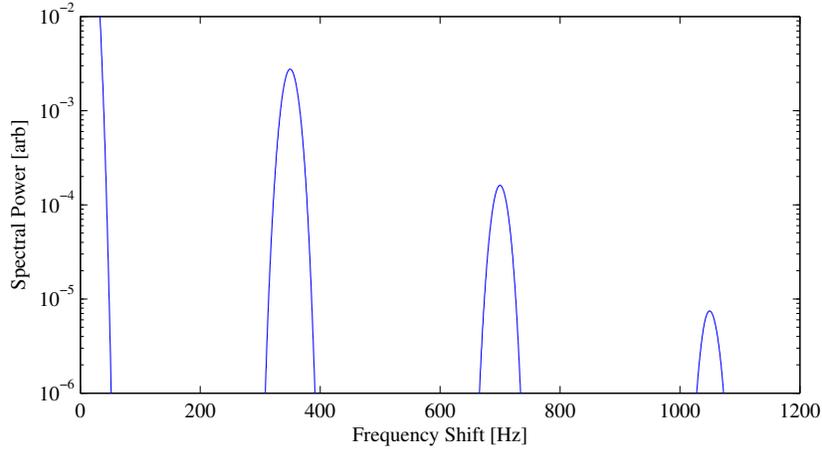}
\caption{\label{fig.pixelspec}Spectral power of the amplitude modulated signal power at pixel (31,32) for the pump power of 0.645. The widths of the peaks are an artifact of the Fourier transform.}
\end{figure}

Otto {\it et al.}\cite{Otto2} also reports measured power modulation of the signal at various positions in the output beam. They transformed the amplitude modulated power to frequency and plotted the spectral powers. We have computed the same type of spectrum at one location near the beam center where both LP$_{01}$ and LP$_{11}$ are strong. The result is shown in Fig. \ref{fig.pixelspec}. Like Otto {\it et al.}\cite{Otto2} we find there are at least three harmonics of 350 Hz with diminishing powers. For other pixels the same three lines appear in the simulated spectra but the relative heights vary.

\subsection{Modal spectra}

We also spectrally resolve the individual fiber modes. As mentioned earlier, for the baseline case in which there is no modulation of the input pump or signal, the sub threshold modal spectra consist of a single unshifted spectral line for LP$_{01}$ and a single Stokes shifted line for LP$_{11}$, matching the signal seed spectra. Well above threshold of the baseline amplifier, the LP$_{11}$ mode begins to amplify a doubly Stokes shifted LP$_{01}$ mode, plus other modes such as LP$_{21}$, also at the doubly shifted frequency. At such high powers the spectrum of LP$_{01}$ consists of lines with Stokes shifts of (0,2)$\Delta \nu$ while the spectrum of LP$_{11}$ has lines at (1,3)$\Delta \nu$. At still higher powers the spectra add more lines with the two modes populating alternating frequencies.

\begin{figure}[htb]
\centering\includegraphics[width=11cm]{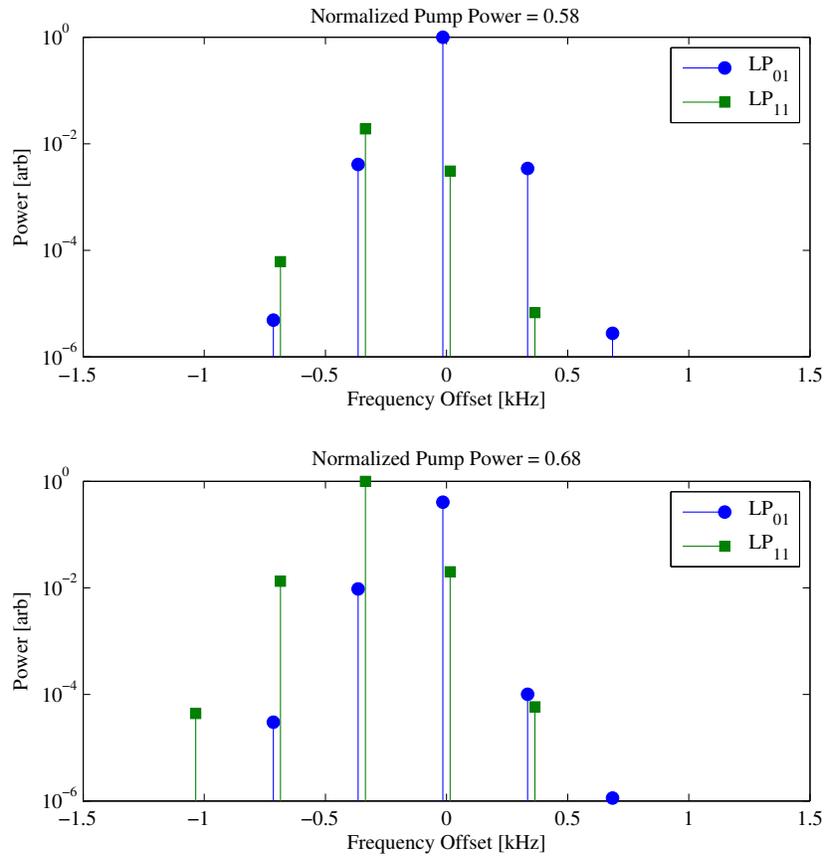}
\caption{\label{fig.modalspec}Modal spectra near threshold (upper plot) and well above threshold (lower plot). The frequencies are slightly shifted for clear display. The conditions are the same as in Fig. 1.}
\end{figure}

The spectra are more complex when the modulated pump is used. Figure \ref{fig.modalspec} shows modal spectra at pump powers 0.58 and 0.68. Well below threshold (not shown) both modes have weak, symmetric sidebands. Slightly above threshold, at pump power 0.58, the Stokes sidebands of LP$_{11}$ are amplified relative to the unshifted light, as shown in the upper plot of Fig. \ref{fig.modalspec}. The center of weight of the LP$_{11}$ spectrum is Stokes shifted relative to that of the nearly centered LP$_{01}$ spectrum at this power. At the higher pump level of 0.68, shown in the lower plot, the LP$_{11}$ light is beginning to convert back to LP$_{01}$ with an additional Stokes shift, and an evolving red shift of LP$_{01}$ becomes apparent.

\section{Conclusions}

We think several of the observed features of amplifier performance reported in ref. \cite{Otto2} suggest that their amplifier is pumped or seeded by modulated light, although we have no direct information about this. We have modeled a similar amplifier design with the assumption of a slightly amplitude modulated pump. This produces several changes compared with the baseline case in which the same amplifier in pumped with unmodulated light and with mode LP$_{11}$ seeded by Stokes shifted light at the quantum noise level. The addition of pump modulation induces amplitude modulation of the individual seeded modes which in turn produces frequency side bands on all modes, making the modal spectra more complex. Above threshold the modal amplitude modulation becomes quite strong, in contrast to the unmodulated pump which produces much weaker modal modulation. 

Our simulation results appear to agree qualitatively with all of the key observations of \cite{Otto2} if we assume that the pump is slightly amplitude modulated at 350 Hz. Modulation of the signal input would produce qualitatively similar behavior. We believe this qualitative agreement between model and measurement strongly supports the mode coupling mechanism we have proposed and implemented in our numerical model. Examination of the laboratory pump light and signal seed light for evidence of amplitude modulation would provide a further critical test.

Finally, we note that we make several approximations in our model, including use of a step refractive index in place of the photonic crystal light guiding structure, co-pumping rather than counter-pumping, symmetric cooling with all four sides of our square spatial grid held at a fixed temperature instead of one-sided cooling. The one-sided cooling used in the experiment is roughly equivalent to a slight bending of the fiber which would displace the modes away from the cooled side. We also include only one orientation of the LP$_{11}$ mode. All of these approximations can be eliminated in future versions of the model if more experimental details are known. Our next task is to speed up the model to allow modeling of counter-pumped fiber with modulated pump or signal.

\section*{Acknowledgments}

This work supported under funding from the Air Force Research Laboratory Directed Energy Directorate.

\end{document}